\documentclass[twocolumn,showpacs,prl,aps]{revtex4}
\usepackage{amssymb}
\usepackage{amsmath}
\usepackage{amsfonts}
\usepackage{graphics}
\usepackage{epic}
\usepackage{eepic}
\usepackage{color}
\usepackage{epsfig}

\begin{document}

\def\ra{\rangle}
\def\la{\langle}
\def\bege{\begin{equation}}
\def\ende{\end{equation}}
\def\begarr{\begin{eqnarray}}
\def\endarr{\end{eqnarray}}
\def\ha{{\hat a}}
\def\hb{{\hat b}}
\def\hu{{\hat u}}
\def\hv{{\hat v}}
\def\hc{{\hat c}}
\def\hd{{\hat d}}
\def\no{\noindent}\def\non{\nonumber}
\def\hi{\hangindent=45pt}
\def\v{\vskip 12pt}

\newcommand{\bra}[1]{\left\langle #1 \right\vert}
\newcommand{\ket}[1]{\left\vert #1 \right\rangle}

\title{An All Linear Optical Quantum Memory 
Based on Quantum Error Correction}

\author{Robert M.\ Gingrich} \email{Robert.M.Gingrich@jpl.nasa.gov}
\author{Pieter Kok} 
\author{Hwang Lee} 
\author{Farrokh Vatan} 
\author{Jonathan P.\ Dowling} 

\affiliation{Quantum Computing Technologies Group, 
Section 367, Jet Propulsion Laboratory, \\
California Institute of Technology, MS 126-347,
 4800 Oak Grove Drive, CA~91109-8099}

\date{\today}


\begin{abstract}
 When photons are sent through a fiber as part of a quantum
 communication protocol, the error that is most difficult to correct
 is photon loss. Here, we propose and analyze a two-to-four qubit encoding
 scheme, which can recover the loss of one qubit in the transmission.
 This device acts as a repeater when
 it is placed in series to cover a distance larger than the
 attenuation length of the fiber, and it acts as an optical quantum
 memory when it is inserted in a fiber loop.
 We call this dual-purpose device
 a ``quantum transponder.'' 
\end{abstract}

\maketitle


Storing qubits for indefinitely long periods of time is a
critically important task in quantum information processing. 
It is needed whenever the quantum
teleportation protocols or feed-forward mechanisms are invoked. 
While photons are ideally suited for the transmission of quantum
information, storing them is very difficult. Currently, techniques of
mapping quantum information between photons and atomic systems are 
being developed \cite{liu01,phillips01,schori02}. 
Obviously, the simplest quantum memory device for photons can be 
an optical-fiber loop or a ring cavity. 
Recently, Pittman and Franson used a Sagnac
interferometer to develop a quantum memory device for photons, robust
to dephasing \cite{pittman02}. 
They concluded that it is, however, the photon loss that
typically restricts the storage time. 

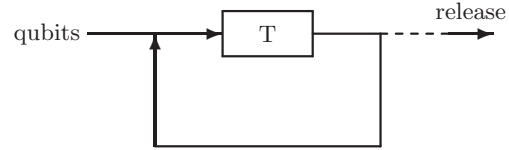
\begin{figure}[t]
\thicklines
\begin{center} 
\unitlength=.3mm 
\begin{picture}(180,60)(0,0)
\drawline(60,40)(60,60)(100,60)(100,40)(60,40)
\drawline(100,50)(130,50)(130,0)(30,0)
\put(30,0){\vector(0,1){50}}
\put(0,50){\vector(1,0){60}}
\dashline[90]{4}(130,50)(160,50)
\put(160,50){\vector(1,0){20}}
\put(80,50){\makebox(0,0){T}}
\put(-18,50){\makebox(0,0){qubits}}
\put(170,60){\makebox(0,0){release}}
\end{picture} \end{center} 

\caption{A cyclic quantum memory using a quantum transponder (T)
based on quantum error correction. 
Placed in series these devices act as
simple quantum repeaters.}
\label{fig:c-memory}
\end{figure}

In this Letter, we present a cyclic quantum memory for photons that
can also deal with photon loss. The idea is to use a delay-line loop
endowed with a small linear-optical quantum computing (LOQC) circuit
\cite{knill01,franson02}, that runs an error-correction code 
over and over again on the loop, as depicted in Fig.~\ref{fig:c-memory}. 
When multiple error-correcting circuits are placed in series, 
it allows the transmission of photons over a distance 
larger than the attenuation length
of the fiber. 
A quantum repeater is a device that executes quantum purification
and swapping protocols--with the goal of achieving remote,
shared, entanglement \cite{briegel98,kok02b,jacobs02}.
Here, in contrast, we define a ``quantum transponder'' to be 
a simple device--one which employs error correction to relay
an unknown quantum state with high fidelity down a quantum channel.
For quantum key distribution schemes such 
as BB84 \cite{bennett84}--only a transponder
is required for long-distance key transfer.
However, we note that if the fidelity of the transponder is 
sufficiently high, we can also use it to distribute entanglement 
by relaying, say, one half of an entangled pair.

\begin{figure}[b]
\thicklines
\begin{center} 
\unitlength=.3mm 
\begin{picture}(230,100)(0,0)
\drawline(0,0)(30,0)
\drawline(30,-10)(30,10)(50,10)(50,-10)(30,-10)
\drawline(50,0)(230,0)
\drawline(0,30)(230,30)
\drawline(0,60)(230,60)
\drawline(0,90)(230,90)
\put(80,0){\circle*{5}}
\put(110,0){\circle*{5}}
\put(140,0){\circle*{5}}
\put(170,90){\circle*{5}}
\put(200,60){\circle*{5}}
\put(80,30){\circle{16}}
\put(110,60){\circle{16}}
\put(140,90){\circle{16}}
\put(170,30){\circle{16}}
\put(200,30){\circle{16}}
\drawline(80,0)(80,38)
\drawline(110,0)(110,68)
\drawline(140,0)(140,98)
\drawline(170,90)(170,22)
\drawline(200,60)(200,22)
\put(40,0){\makebox(0,0){$H$}}
\put(-8,0){\makebox(0,0){$\ket{0}$}}
\put(-8,30){\makebox(0,0){$\ket{0}$}}
\put(-10,60){\makebox(0,0){$\ket{q_2}$}}
\put(-10,90){\makebox(0,0){$\ket{q_1}$}}
\end{picture} \end{center} 

\caption{Two-to-four qubit encoding, which
converts the input $|q_1, q_2\ra$ into a four-qubit state, 
as is given in Eq.~(\ref{code-equ}).
}
\label{fig:ecc} 
\end{figure}
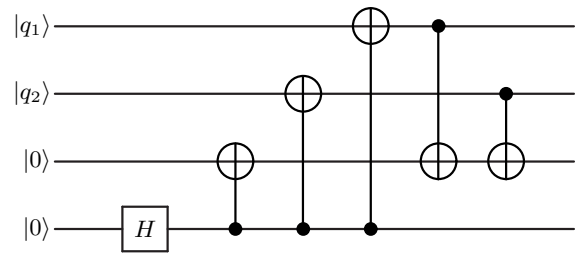

Let us first consider our error-correcting code.
We use a two-to-four one-error--correcting scheme for protecting the data
against photon loss. That is, we encode two qubits into four qubits,
such that the resulting code is capable of recovering from the loss of
one photon. The encoding is as follows: 
\begin{equation}
 \begin{array}{rcl}
 \ket{00} & \mapsto & (\ket{0000}+\ket{1111})/\sqrt{2}\; , \\
 \ket{01} & \mapsto & (\ket{0110}+\ket{1001})/\sqrt{2}\; , \\
 \ket{10} & \mapsto & (\ket{1010}+\ket{0101})/\sqrt{2}\; , \\
 \ket{11} & \mapsto & (\ket{1100}+\ket{0011})/\sqrt{2}\; .
\end{array} 
\label{code-equ}
\end{equation}
This code was first introduced in 1997 by Grassl, Beth and Pellizzari
\cite{grassl}, and it can be implemented using the simple quantum
circuit shown in Fig.~\ref{fig:ecc}.

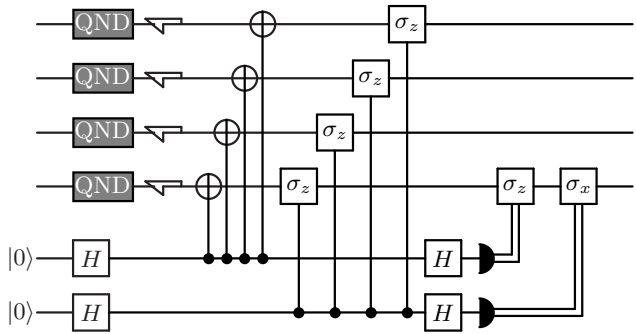
\begin{figure}[bt]

\thicklines
\begin{center} 
\unitlength=.24mm 
\begin{picture}(330,180)(-7,-60)
\drawline(0,0)(20,0)
\put(20,-8){\shade\path(0,0)(33,0)(33,16)(0,16)(0,0)}
\put(36,0){\makebox(0,0){{\color{white}\small QND}}}
\drawline(53,0)(65,0)
\drawline(60,3)(80,3)
\drawline(60,3)(70,-5)
\drawline(70,-5)(70,0)
\drawline(70,0)(135,0)
\drawline(80,0)(80,3)

\drawline(0,30)(20,30)
\put(20,22){\shade\path(0,0)(33,0)(33,16)(0,16)(0,0)}
\put(36,30){\makebox(0,0){{\color{white}\small QND}}}
\drawline(53,30)(65,30)
\drawline(60,33)(80,33)
\drawline(60,33)(70,25)
\drawline(70,25)(70,30)
\drawline(70,30)(155,30)
\drawline(80,30)(80,33)

\drawline(0,60)(20,60)
\put(20,52){\shade\path(0,0)(33,0)(33,16)(0,16)(0,0)}
\put(36,60){\makebox(0,0){{\color{white}\small QND}}}
\drawline(53,60)(65,60)
\drawline(60,63)(80,63)
\drawline(60,63)(70,55)
\drawline(70,55)(70,60)
\drawline(70,60)(175,60)
\drawline(80,60)(80,63)

\drawline(0,90)(20,90)
\put(20,82){\shade\path(0,0)(33,0)(33,16)(0,16)(0,0)}
\put(36,90){\makebox(0,0){{\color{white}\small QND}}}
\drawline(53,90)(65,90)
\drawline(60,93)(80,93)
\drawline(60,93)(70,85)
\drawline(70,85)(70,90)
\drawline(70,90)(195,90)
\drawline(80,90)(80,93)

\drawline(0,-40)(20,-40)
\drawline(20,-50)(20,-30)(40,-30)(40,-50)(20,-50)
\drawline(0,-70)(20,-70)
\drawline(20,-80)(20,-60)(40,-60)(40,-80)(20,-80)
\put(30,-40){\makebox(0,0){$H$}}
\put(30,-70){\makebox(0,0){$H$}}
\put(-9,-40){\makebox(0,0){$\ket{0}$}}
\put(-9,-70){\makebox(0,0){$\ket{0}$}}
\put(95,0){\circle{14}}
\put(105,30){\circle{14}}
\put(115,60){\circle{14}}
\put(125,90){\circle{14}}
\put(95,-40){\circle*{5}}
\put(105,-40){\circle*{5}}
\put(115,-40){\circle*{5}}
\put(125,-40){\circle*{5}}
\drawline(95,-40)(95,7)
\drawline(105,-40)(105,37)
\drawline(115,-40)(115,67)
\drawline(125,-40)(125,97)
\drawline(135,-10)(135,10)(155,10)(155,-10)(135,-10)
\drawline(155,20)(155,40)(175,40)(175,20)(155,20)
\drawline(175,50)(175,70)(195,70)(195,50)(175,50)
\drawline(195,80)(195,100)(215,100)(215,80)(195,80)
\put(145,-70){\circle*{5}}\drawline(145,-70)(145,-10)
\put(165,-70){\circle*{5}}\drawline(165,-70)(165,20)
\put(185,-70){\circle*{5}}\drawline(185,-70)(185,50)
\put(205,-70){\circle*{5}}\drawline(205,-70)(205,80)
\put(145,0){\makebox(0,0){$\sigma_z$}}
\put(165,30){\makebox(0,0){$\sigma_z$}}
\put(185,60){\makebox(0,0){$\sigma_z$}}
\put(205,90){\makebox(0,0){$\sigma_z$}}
\drawline(215,-50)(215,-30)(235,-30)(235,-50)(215,-50)
\drawline(215,-80)(215,-60)(235,-60)(235,-80)(215,-80)
\put(225,-40){\makebox(0,0){$H$}}
\put(225,-70){\makebox(0,0){$H$}}
\drawline(40,-40)(215,-40)
\drawline(40,-70)(215,-70)

\put(245,-40){\blacken\arc{16}{-1.57}{1.57}}
\put(245,-70){\blacken\arc{16}{-1.57}{1.57}}
\drawline(235,-40)(245,-40)
\drawline(235,-70)(245,-70)
\drawline(250,-38)(263,-38)
\drawline(250,-42)(267,-42)
\drawline(255,-10)(255,10)(275,10)(275,-10)(255,-10)
\put(265,0){\makebox(0,0){$\sigma_z$}}
\drawline(263,-38)(263,-10)
\drawline(267,-42)(267,-10)

\drawline(290,-10)(290,10)(310,10)(310,-10)(290,-10)
\put(300,0){\makebox(0,0){$\sigma_x$}}
\drawline(250,-68)(298,-68)
\drawline(250,-72)(302,-72)
\drawline(298,-68)(298,-10)
\drawline(302,-72)(302,-10)
\drawline(155,0)(255,0)\drawline(275,0)(290,0)\drawline(310,0)(330,0)
\drawline(175,30)(330,30)\drawline(195,60)(330,60)\drawline(215,90)(330,90)

\end{picture} \end{center} 
\caption{Quantum transponder that
recovers photon loss (here, for example, in the lower-most qubit) 
using two ancilla photons.
The QND box represents a single-photon quantum nondemolition measurement
device, followed by a single-photon source depicted by the 
gun-shaped polygon.
$H$ represents the Hadamard gate.
Four CNOT (controlled by the first ancilla) 
and four CZ (controlled by the second ancilla) gates are followed
by another Hadamard gate and measurement on the computational basis
for each ancilla.
The final one-qubit operations are for the channel where
the loss has occurred, and depends on the measurement results, see
TABLE I.
It is depicted here for loss in the lower-most channel.}
\label{fig:recovery} 
\end{figure}

The difference between this code, and the usual error-correcting codes
based on syndrome detection, is that we do not destroy the
ancill\ae. In other words, we recover from a single-photon loss without
losing any of the four qubits. 
The error-correction process is shown in Fig.~\ref{fig:recovery}, 
where we assumed that the loss has
occurred in the lower-most qubit (the conditional error-correcting
operator consisting of a combination of $\sigma_x$ and $\sigma_z$ acts
on the lower-most qubit mode; here, $\sigma_x$, $\sigma_y$, and $\sigma_z$
are the Pauli matrices). Similar circuits work for photon loss in the
other three modes.

We can demonstrate how this algorithm works by studying its action on
one of the codewords of the code of Eq.~(\ref{code-equ}). 
For example,
consider the codeword $\ket{\psi_0}=\frac{1}{\sqrt{2}} \left(
  \ket{0110}+\ket{1001} \right)$, and suppose that the last qubit is
lost (in accordance with Fig.~\ref{fig:recovery}). The state of the
system is given by the following density operator $\rho_1 =
{\textstyle \frac{1}{2}} ( \ket{011}\bra{011} + \ket{100}\bra{100} )$, 
which is obtained from the initial state $\ket{\psi_0}$ by tracing-out
the last qubit. In what follows, it is easier to consider the mixed
state $\rho_1$ as a probability distribution over the pure states,
instead of a density matrix. Thus the mixed state after photon loss
can be written as $\rho_1 = \{ (\ket{011},{\textstyle \frac{1}{2}}),
(\ket{100},{\textstyle \frac{1}{2}}) \}$. 

The quantum nondemolition
(QND) device that signals the loss of the last qubit is followed by a
qubit state preparation device (photon gun) 
that substitutes the missing qubit with
a new qubit in the ground state $\ket{0}$. 
The new density operator is then:
$\rho_2 = \{ (\ket{0110},{\textstyle \frac{1}{2}}), (\ket{1000},
{\textstyle \frac{1}{2}}) \}$.  
Including the two ancilla bits, the total system is in the mixed state
$\rho_3=\rho_2 \otimes \ket{00}\bra{00} = \{ (\ket{0110} \ket{00},
{\textstyle \frac{1}{2}}), (\ket{1000} \ket{00}, {\textstyle
  \frac{1}{2}}) \}$. After applying the Hadamard transform on the
ancilla bits, this becomes 
\begin{multline}
 \rho_4 = \big\{ 
  \big( {\textstyle
  \frac{1}{2}}\ket{0110}(\ket{00}+\ket{01}+\ket{10}+\ket{11}), 
     {\textstyle \frac{1}{2}} \big), \\
  \big( {\textstyle
  \frac{1}{2}}\ket{1000}(\ket{00}+\ket{01}+\ket{10}+\ket{11}), 
     {\textstyle \frac{1}{2}}) \big) \big\} . 
\end{multline} 
                           
The four controlled-$\sigma_x$ (CNOT) and controlled-$\sigma_z$ (CZ)
operations, followed by the the Hadamard transform on the ancilla
bits, then yields the mixed state
$\rho_5= 
  \big\{ \big(
    {\textstyle \frac{1}{2}} \big( (\ket{0110}+\ket{1001})\ket{00}+
    (\ket{0110}-\ket{1001})\ket{10} \big), {\textstyle \frac{1}{2}} \big), \\
   \big(
    {\textstyle \frac{1}{2}} \big( (\ket{1000}+\ket{0111})\ket{01}+
    (\ket{1000}-\ket{0111})\ket{11} \big), {\textstyle \frac{1}{2}} \big)\big\}.
$

Finally, the measurement outcome of the two ancill{\ae} 
determines the
error-correcting operator on the last qubit. These conditional
operators are listed in Table~\ref{tab:operators}. 
Note that the after
the measurement of the ancill\ae, the result is always a {\em pure}
state. Furthermore, all the results are equally likely, so this
process does not reveal any information about the original encoded
state. It follows immediately from the Table I and
Eq.~(\ref{code-equ}) that this will correct the loss of the
photon for this particular codeword.  In a similar fashion,
it can be shown that the
error correction will work for an arbitrary input state and a single
lost photon.  

\begin{table}[t]
\caption{\label{tab:operators} The measurement outcomes of the
  ancill\ae ~and the conditional error-correcting operators that
  restore the state of the encoded qubits.}
\begin{ruledtabular}
\begin{tabular}{ccc}
observation & projected state & correcting operation \\ \hline
$\ket{00}$ & $\ \ket{0110}+\ket{1001}$ & $I$ \\
$\ket{01}$ & $\ \ket{1000}+\ket{0111}$ & $\sigma_x$ \\
$\ket{10}$ & $\ \ket{0110}-\ket{1001}$ & $\sigma_z$ \\
$\ket{11}$ & $\ \ket{1000}-\ket{0111}$ & $\sigma_x\, \sigma_z$
\end{tabular}
\end{ruledtabular}
\end{table}

In the remainder of this paper we will consider optical
implementations of this error-correcting code, where the qubits are
encoded in two-mode single-photon states. These may either be two
spatial modes 
or two polarization modes \cite{knill01,koashi01,franson01,pittman02a}.

When we send a photon through an optical fiber of length $d$, the
probability of successfully transmitting the photon is given by
\begin{equation}
\label{pdef}
  p(d) = \exp(- \alpha d)\; .
\end{equation}
Here, the absorption coefficent of the fiber is given by $\alpha$, 
which is a property of the fiber. 
The best fibers have an absorbtion length, $1/\alpha$,
of about 30 km.  We wish to overcome this length restriction with the
error correcting codes (ECC) and the linear-optical scheme for
implementing two qubit gates that was introduced by Knill, Laflamme,
and Milburn (KLM)~\cite{knill01}.

The ECC described above can recover two qubits, given the loss of a single
photon. With a perfectly working ECC, the probability of losing zero
or one photon in the fiber over a distance
$d$ is given by   
$
p_f = p^4 + 
4 p^3 (1 - p) \; ,
$
where $p$ is given by Eq.~(\ref{pdef}).  Using $p_f$ and inverting
equation~(\ref{pdef}) we can calculate an effective absorbtion length
for the ECC, or equivalently $\alpha'$:
\begin{eqnarray}
  \alpha'(\alpha,d) & = & - \ln (p_f) /d \nonumber \\
                      & = & 3 \alpha - \ln (4 - 3 \exp (- \alpha
                      d)) /{d}  \; .
\end{eqnarray}
Since our ECC encodes two qubits, we compare $\alpha'$ with $2 \alpha$,
to see if our code is improving the situation or not.  Define the
function $f(x)$ with $x \equiv \alpha d$, such that 
\begin{eqnarray}
  \frac{\alpha'}{2 \alpha} = \frac{3}{2} - \frac{\ln ( 4 - 3 \exp
       (-x))}{2 x} \equiv f(x)\; .
\end{eqnarray}
When $x < \ln (3) \approx 1.1$, $f(x) < 1$, 
our ECC (transponder) is increasing the effective 
absorption length for the qubits
we are trying to transmit.  So, if we make $d < \ln (3) / \alpha$ the
transponder allows us to transmit qubits with higher fidelity than is possible
without it.  
Note that $\lim_{x \rightarrow 0} f(x) = 0$, so the
absorbtion length can be made arbitrarily large by making $d$ smaller.  
However, by decreasing $d$ we need to introduce more gates, and the
gates introduce errors.  

We employ the KLM scheme to implement our quantum circuit. As a
consequence, all one-qubit gates can be implemented with minimal errors. 
Furthermore, we can execute a controlled-not
(CNOT) or controlled-sign (CZ) operation efficiently by using ancilla
qubits.  For $2n$ ancill\ae, a CNOT or CZ gate can be
successfully executed with probability
\begin{equation}
\label{gateprob}
  p = \left( \frac{n}{n + 1} \right)^2 , \;
\end{equation}
assuming all single photon guns, QND measurements, and photon
detections work perfectly.

Note that when we send our qubit through a fiber, we are 
encoding the information
in the polarization of a photon, i.e.,
$  | H \rangle \rightarrow | 0 \rangle \; \; \; | V \rangle \rightarrow |
    1 \rangle \; 
$.
In the ``dual-rail'' KLM scheme, the qubit is encoded in 
the path (upper or lower) of the photon.  We are therefore sending our qubits in the
polarization basis but we are doing error correction in the position
basis.  This is not a problem because we can use a polarizing beam
splitter and appropriate polarization rotators to convert between the
two bases. In particular, every KLM-based gate has an equivalent
implementation for polarization-encoded qubits.

\begin{figure}[t]
\centerline{\psfig{figure=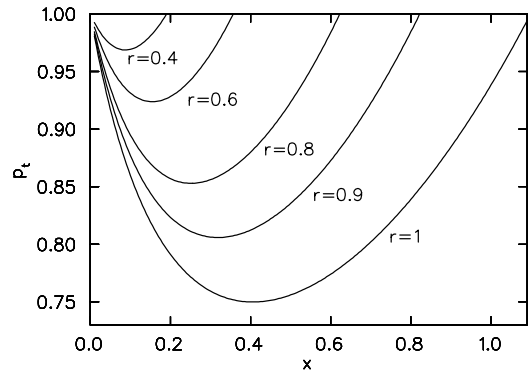,width=4.8cm,angle=90}}
\bigskip
\caption{\label{contour} A contour plot of $r = \alpha' (x,n)/2
    \alpha$, the relative absorption coefficient,
    as a function of $x$, the normalized distance, and
    the success probability of the transponder, $p_t$. 
    Note that the minimum of the $r = 1$ curve is at $p_t =
    {3}/{4}$.  }
\end{figure}

With imperfect gates, our equation for $r\equiv \alpha' / 2 \alpha$ becomes
\begin{eqnarray}
\label{req}
 r \equiv \frac{\alpha' (x,n)}{2 \alpha} & = & - \frac{1}{2 x} 
\ln \left(
 p_f p_t \right) \cr
  & = & f(x) + \frac{1}{2 x} \ln \left( \frac{1}{p_t} \right) \;,
\end{eqnarray}
where $p_t$ is the probability that all the gates in the 
quantum transponder work correctly.  
We can see that for $p_t < 1$, the minimum of 
$r$ is no longer at $x = 0$ (where the transponder stations are
placed back-to-back), since the second term is infinite at that point.  
Figure~\ref{contour} is a contour plot of Eq. (\ref{req})
as a function of $x$ and $p_t$.
Note that the losses from
the gates that do the encoding and decoding of the qubits are one-time
losses, and become relatively unimportant for long transmission lines.

Using Eq.~(\ref{gateprob}), and noting that there are four CNOT 
and four CZ gates in each transponder, we can write
$p_t=[n/(n+1)]^{16}$. 
The value of $n$ for which the minimum drops below one is at $n = 56$.
So we need at least 112 ancilla qubits at each gate for the transponder to
transmit qubits more reliably than the fiber without error correction.

Let us now consider the probability of success for a single quantum
transponder with inefficient detectors. 
Suppose, as in Fig.~\ref{fig:recovery}, we use the QND device
proposed by Kok {\em et al}.\ \cite{kok02}. 
This device operates by teleporting
the input photon to the output mode, using coincidence counting in a
CNOT-operated Bell measurement. The photon loss is then signalled by
finding a single detector click.
In this case, there is always a photon in the output mode, 
and we do not need the four single-photon guns.

As shown in Fig.~\ref{fig:recovery}, the transponder
consists of two single-photon guns (SPG), four QND devices, six
one-qubit gates, four CNOT gates, four CZ gates, and two
photodetectors (see the first row in Table~\ref{table}). 
The single-photon QND measurement can be accomplished with two SPGs, 
two CNOT gates, two Hadamard gates, and two photodetectors.
The first Hadamard and CNOT gates are for Bell state preparation, 
and the Bell state measurement can be made by CNOT and Hadamard and
measurements in the computational basis. For each CNOT gate, we have
two one-qubit gates and a CZ gate. 

\begin{table}[b]
\caption{\label{table} Number of single-photon guns, QND devices,
  CNOT, CZ, one-qubit gates, and photodetectors per
  transponder. (i) Each QND device
  consists of two SPG, two CNOT, and two photodetectors. (ii) Each
  CNOT can be considered as a CZ and two one-qubit gate. (iii)  Each
  CZ requires $2n$ ancilla photons and $2(n+1)$ photodetectors (PD).}
\begin{ruledtabular}
\begin{tabular}{ccccccc}
  ECC   & SPG        & QND & CNOT & CZ & One   & PD        \\ \hline
  \quad & 2          & 4   & 4    & 4  & 6     & 2           \\
  (i)   & 2+8        & 0   & 4+8  & 4  & 6+8   & 2+8           \\
  (ii)  & 10         & 0   & 0    & 16 & 14+24 & 10              \\
  (iii) & $10 + 32n$ & 0   & 0    & 16 & 38    & $10 + 32(n+1)$ 
\end{tabular}
\end{ruledtabular}
\end{table}

Given that we use $2n$ ancilla photons for each CZ gate, we need to
have $2n$ SPGs and $2(n + 1)$ photodetectors. Altogether, we need to
have 38 one-qubit gates, sixteen CZ gates, $10 + 32n$ SPGs, and $10 +
32(n+1)$ photo detectors. Hence the probability of success for the
transponder is given by
$
  p_{t} = p_{\rm one}^{38} ~p_{\rm two}^{16} 
          ~p_{\rm spg}^{10+32n} ~\eta^{10+32n}\; ,
$
where $p_{\rm one}$, $p_{\rm two}$, and $p_{\rm spg}$ are the success
probability of the QND measurement, the one-qubit gate, the two-qubit
(CZ) gate, and a SPG, respectively. Note that $\eta$ denotes the
quantum efficiency of the photo detector, where $10+32n$ detectors
among $10+32(n+1)$ should click for perfect gate operations. 

\begin{figure}[t]
\centerline{\psfig{figure=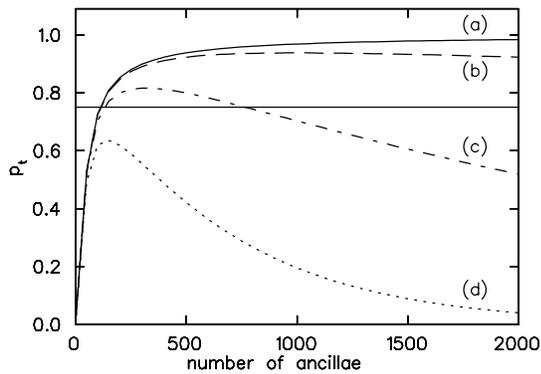,width=4.8cm,angle=90}}
\bigskip
\caption{\label{fig:efficiency} The probability of transponder success 
   $p_t$ as a function of the number of ancill{\ae} $n$. 
    The top graph, (a) is for detector efficiency $\eta=1$, 
    and the descending graphs
    have detector efficiencies (b) $1-10^{-6}$, (c) $1-10^{-5}$, and
    (d) $1-10^{-4.5}$, respectively. A typical value that $p_{t}$
    needs to exceed is 0.75 (the dashed line).} 
\end{figure}

Now let us assume that the number of ancilla photons used for a
two-qubit gate is $2n$, which gives $p_{\rm two}=n^2 /
(n+1)^2$. Hence, the number of ancilla photons may be optimized for a
given quantum efficiency of the photodetectors and the success
probability of the single photon guns. Let us assume for now that 
$p_{\rm one}= p_{\rm spg}=1$. Then $p_{t}$ is given by 
\begin{equation}
  p_{t} = p_{\rm two}^{16} ~\eta^{10+32n}
  =\left( n \over n+1\right)^{32} \eta^{10+32n}\; . 
\end{equation}
For example, if $1-\eta=10^{-5}$, taking $n=16$ yields 
$p_t \approx 0.14$. With $n=160$ we have $p_{t} \approx 0.78$. A
typical value that the success probability needs to beat is 0.75. 
In Fig.~\ref{fig:efficiency} we plot $p_{t}$ as a function of
the number of ancill{\ae} with different detector efficiencies $\eta$.

In conclusion, we have presented an error-correction scheme that
encodes an unknown two photon state into four photons, up to one of which 
can be lost in the transmission. 
This device acts as a simple repeater or quantum transponder 
when it is placed in series, and it acts
as an optical quantum memory when it is inserted in an optical loop.
Since the absorbtion length for two photons in a fiber is $1/(2
\alpha)$, the storage time is given by $T_f = 1/(2 \alpha \nu)$ where
$\nu$ is the speed of light in the fiber.  With error correction we
can increase the storage time to $T_f / r$.
We gave a quantitative analysis of the behaviour of 
this quantum memory 
in several situations, deriving values for the optimal
length of the loop,  
and characterizing the performance in the presence of detector losses.

Using this scheme, the conversion between flying qubits and stationary
qubits in memory is not necessary with LOQC, as the memory and quantum
logic gates are composed of the same optical resources. 
The delay line, when rolled out, is a fiber quantum communication line 
with simple LOQC transponders, suitable for 
the BB84 quantum key distribution protocol \cite{bennett84}. 



This work was carried out at the Jet Propulsion Laboratory, California
Institute of Technology, under a contract with the National Aeronautics 
and Space Administration. We wish to thank J.D.\ Franson, G.J.\
Milburn, and T.B.\ Pittman for stimulating discussions, and we would
like to acknowledge support from the National Security Agency, the
Advanced Research and Development Activity, the Defense Advanced
Research Projects Agency, the National Reconnaissance Office, and the
Office of Naval Research. R.M.G.\ and P.K.\ thank the National
Research Council and NASA Code Y for support.

\end{document}